\documentclass[useAMS,usenatbib]{mn2e}
\usepackage{graphics}

\title[Large scale circulations and energy transport in contact binaries]
{Large scale circulations and energy transport in contact binaries}

\author[K. St\c{e}pie\'n]
{K. St\c{e}pie\'n$^{1}$\thanks{e-mail: kst@astrouw.edu.pl} \\
$^{1}$Warsaw University Observatory, Al. Ujazdowskie 4, 00-478 Warszawa, Poland}

\date{Accepted --.
         Received -- ;
         in original form --}

\pubyear{2007}
\begin{document}
\maketitle
\label{firstpage}
\begin{abstract}
A hydrodynamic model for the energy transport between the components of a
contact binary is presented. Energy is transported by a large-scale, steady
circulation carrying high entropy matter from the primary to secondary
component. The circulation is driven by the baroclinic structure of the
common envelope, which is a direct consequence of the nonuniform heating at
the inner critical Roche lobes due to unequal emergent energy fluxes of the
components. The mass stream flowing around the secondary is bound to the
equatorial region by the Coriolis force and its width is determined
primarily by the flow velocity. Its bottom is separated from the underlying
secondary's convection zone by a radiative transition layer acting as an
insulator. For a typically observed degree of contact the heat capacity of
the stream matter is much larger than radiative losses during its flow
around the secondary. As a result, its effective temperature and entropy
decrease very little before it returns to the primary. The existence of the
stream changes insignificantly specific entropies of both convective
envelopes and sizes of the components. Substantial oversize of the secondaries,
required by the Roche geometry, cannot be explained in this way. The
situation can, however, be explained by assuming that the primary is a main
sequence star whereas the secondary is in an advanced evolutionary stage with
hydrogen depleted in its core. Such a configuration is reached past mass
transfer with mass ratio reversal. Good agreement with observations is
demonstrated by model calculations applied to actual W UMa-type
binaries. In particular, a presence of the equatorial bulge moving with a
relative velocity of 10-30 kms$^{-!}$ around both components of AW UMa is
accounted for.

\end{abstract}

\section{Introduction}  
\label{sect:intro}

Contact binaries have been defined by \citet{kui41} as binaries with
components surrounded by a common envelope. Within the Roche approximation
for the total potential in a rotating frame of reference, the envelope lies
between the inner and outer critical equipotential surfaces, defined
respectively by the Lagrangian points L$_1$ and L$_2$. Cool contact
binaries with spectral types later than F0 are called W UMa type binaries
\citep{moch81}.

W UMa type stars are fairly common in space. \citet{ruc02} estimates that
one such binary occurs per 500 main sequence stars in the solar vicinity
but these stars do not appear among members of young and intermediate age
clusters.  Their occurrence rapidly increases in old open and globular
clusters \citep{kr93, ruc98, ruc00}. This result indicates that cool
contact configurations are not formed at, or near the zero age main
sequence (ZAMS) but at the substantially later age. As it is now commonly
assumed, they are formed from initially detached binaries which lose orbital
angular momentum (AM) at such a rate that a contact is reached after several
Gyr. Kinematically, field W UMa stars belong to old disk, which supports their
advanced age \citep{gb88, bil05}.  The most promising mechanism for AM loss
is related to the chromospheric-coronal activity of binary components
\citep{vil82, moch85, ste95} although a high incidence of companions to W
UMa stars \citep{ruc07} suggests that a third body may also play a role in
orbit tightening \citep{egg01}.

\citet{kui41} noted that a contact binary is stable only when both
components are identical. Otherwise, a binary consisting of two stars with
the same composition is unstable and mass will be transferred inside the
common envelope from the more massive (primary) to the less massive
(secondary) component. The transfer is driven by a nonuniform heating of
the base of the common envelope due to unequal emergent luminosities at the
inner critical surface. By continuity, temperature differences on each
equipotential surface above the Roche lobes exist. As a result, the common
envelope must be treated as baroclinic rather than barotropic \citep{shu79,
tas92}. Different vertical (i.e. perpendicular to the local equipotential
surface) pressure stratifications in both components produce the horizontal
pressure gradient driving mass motions on a dynamical time scale. Even if
the pressure is constant over a given equipotential surface, a difference
appears above and below that surface \citep{kah04}. In the absence of the
Coriolis force the resulting large scale flows are symmetric around the
axis joining centers of both stars \citep{web76}. However, for a flow
velocity being a significant fraction of the sound velocity the Coriolis
force cannot be neglected. The force acts outwards in the equatorial
plane, deflects the flow towards one side of the neck between the
components and makes it go around the other star in the direction of its
orbital motion. Figure 1 shows geometry of both critical Roche lobes and of
the flow in the equatorial plane (see also Fig.~4 in
\citealt{kui41}). Subsequent computations of the streamlines in
semidetached binaries confirmed a strong influence of the Coriolis force
(see e.g. \citealt{ls75}, \citealt{ls76}, \citealt{oka02}). When the flow
returns to the parent star it is directed to the other side of the neck
(Fig.~1). 

Together with mass, thermal energy is carried to the other component, so we
should expect a significant modification of the apparent surface brightness
of both stars, compared to a detached binary. Indeed, observations of W UMa
type stars show eclipse minima of nearly equal depth indicating almost the
same surface brightness of both components. It indicates that, unless mass
ratio is close to one, a substantial fraction of the flux radiated
by the secondary component comes from the primary.

\begin{figure}
\begin{center}
\rotatebox{0}{\scalebox{0.48}{\includegraphics{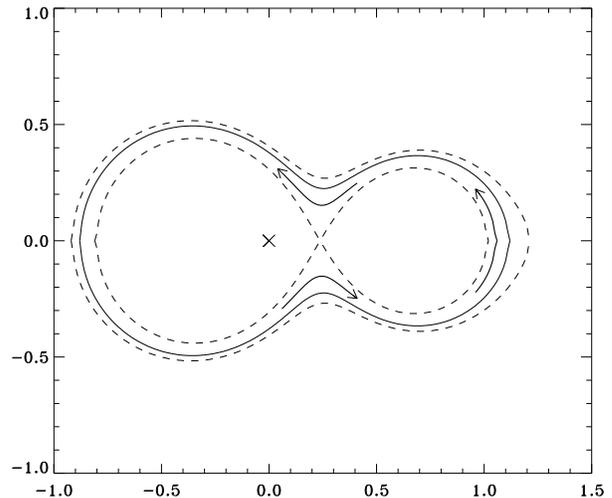}}}
\caption{\label{f1}
Geometry of a contact binary in an equatorial plane. The binary rotates
anticlockwise around the axis crossing the plane in a point marked with a
cross. Broken lines denote inner and outer critical surfaces. A common
envelope lies between the inner critical surface and the stellar surface
plotted with a solid line. Mass flows from the primary through a neck (the
stellar surface narrowing) and encircles the secondary, as indicated by
arrows. (For a side view of a contact binary see Fig.~5).}
\end{center}
\end{figure}

A very elegant model of a W UMa type binary consisting of two main sequence
stars was developed by \citet{lucy68, lucy76} and \citet{flan76} (see also
\citealt{ye05} and references therein). The model assumes that each
component is out of thermal equilibrium, with a radius
oscillating around its critical Roche lobe (inner critical surface). 
The model requires that matter transported
from the primary fully covers the secondary like a hot blanket. The blanket
blocks completely the radiation energy flux produced in the core of the
secondary. The blocked flux is converted into the thermal energy of the
secondary which expands on the stellar thermal time scale. Specific entropy
in its convection envelope increases, approaching the value corresponding
to the primary's convection zone, which, at the same time, decreases
somewhat due to energy transfer. When the expanding secondary overflows its
Roche lobe, mass and energy transfer from the primary is stopped and the
star shrinks radiating away the excess thermal energy until the cycle of
these thermal relaxation oscillations (TRO) repeats. The TRO model
successfully explained two important properties of W UMa type stars: an
abnormal radius ratio of the components (assumed to be MS stars), required
by the Roche geometry, and essentials of the observed light curves.

In this paper, calculations are presented that reveal difficulties
with the TRO model. In particular, it will be shown below that the Coriolis
force is strong enough to balance the meridional pressure
gradient of the circulation current. As a result, the stream of
matter from the primary is bound to the equatorial belt of the secondary.
The polar regions are not covered by the stream so the star can freely
radiate away its nuclear energy. As a result, the specific entropy in its
convection zone increases insignificantly and remains considerably below the
value characteristic of the primary convection zone. Also its radius
hardly increases, which excludes MS secondaries  oversized by a factor of 2-3
observed in many W UMa stars \citep{ste06a}. Substantially oversized
secondaries can, however, be naturally explained by assuming that, instead
of being MS objects, they are highly evolved stars with depleted hydrogen
in the center or even possessing small helium cores \citep{pacz07}.

A possibility of existing cool contact binaries with secondaries in a
more advanced evolutionary stage than primaries has been
mentioned in a number of papers \citep{tw75, sar89, egg96} but such systems
were considered to be exceptions from the general rule stating that W UMa
stars have not reversed mass ratios. An evolutionary model with mass ratio
reversal being a {\em necessary} condition for forming a W UMa type binary
was proposed by \citet{ste04, ste06a, ste06b}. The starting configuration
for such a model is a close detached binary with a initial orbital period
of a couple of days. The binary loses AM via a magnetized wind from both
components. The AM loss rate varies approximately as $M^{-3}$, where $M$ is
a stellar mass \citep{gs08}.  This dependence is very similar to the mass
dependence of the MS evolutionary time scale. Because of this coincidence
the primary is expected to be close to, or just beyond terminal age MS
(TAMS) at the time when the Roche lobe descends onto its surface and
the mass transfer to the secondary begins.  Conservative, or nearly
conservative mass transfer makes the orbit shrink, which accelerates the
transport rate. The process stops after majority of the primary's mass is
transferred and the mass ratio reversed (i.e. the former secondary
component becomes now the primary). Depending on the detailed values of the
binary parameters and the mass transfer process, a contact
binary or a very short period Algol-type star emerges \citep{ste06a}. 
In the latter case
an additional AM loss via the magnetized wind is needed to turn the Algol
into a contact binary. After receiving hydrogen-rich material from the
secondary, the primary moves upward toward ZAMS (corresponding to its
larger mass), whereas the secondary becomes substantially oversized for its
(lower) mass. Each star separately is in thermal equilibrium while filling
its critical Roche lobe. Further evolution of the binary is governed by a
self-regulating mechanism with two processes acting in the opposite
directions: evolutionary expansion of the secondary, followed by mass
transfer to the primary, which leads to widening of the orbit and orbital
AM loss tightening it. As a result, a contact configuration is maintained
with a slow (on a nuclear evolution time scale) net mass transfer from the
secondary to primary \citep{gs08} until coalescence occurs when the mass
ratio reaches a critical value \citep{ras95}.

The model explains the Kuiper paradox, i. e. the existence of an
equilibrium configuration with component sizes required by the Roche
geometry. It does not, however, explain the observed properties of light
curves. The energy transport between the components is not a part
of that model. It is an independent process which will be considered in
detail in the present paper.

The paper is organized as follows: basic assumptions and equations
governing the mass and energy flow are given in Sect. 2. The stream
structure across and along the flow is discussed in Sect. 3, together with
application of the derived formulas to two actual contact
systems. Section 4 describes the global reaction of both components to the
circulation and Sect. 5 discusses the results of the paper. It also
summarizes the main results.

\section{Equations and assumptions}

The basic Eulerian equations of fluid flow in a frame of reference rotating
with the binary are:

\noindent continuity equation

\begin{equation}
\frac{\partial\rho}{\partial t} +
     \mathbf{\nabla\cdot}(\rho\mathbf{v}) = 0\,,
\end{equation}

\noindent momentum equation

%\begin{eqnarray}
\begin{equation}
\frac{\partial \mathbf{v}}{\partial t} + \mathbf{(v\cdot\nabla)v} =
\frac{1}{\rho}\mathbf{\nabla}p + \nu\nabla^2\mathbf{v}
%+ \left(\frac{\zeta}{\rho} +
%\frac{1}{3}\nu\right)\mathbf{\nabla(\nabla\cdot v)}\\
- \mathbf{\nabla}\varphi -2\mathbf{\Omega\times v}\,,%\nonumber
%\end{eqnarray}
\end{equation}

\noindent and entropy equation

\begin{equation}
\rho T\left[\frac{\partial s}{\partial t} + (\mathbf{v\cdot\nabla)}s\right] =
- \mathbf{\nabla\cdot F} + d_{\mathrm{visc}}\,.
\end{equation}

Here $\rho, p, s$ and $\mathbf{v}$ denote gas density, pressure, entropy
and velocity, $\varphi$ is the gravitational (plus centrifugal) potential,
$\nu$ is kinematic viscosity coefficient,
$d_{\mathrm{visc}}$ is the viscous term, $\mathbf{F}$ is the radiative
energy flux and $\Omega$ is angular velocity. In the momentum equation the
viscous term with $\mathbf{\nabla(\nabla\cdot v)}$ has been left out as
unimportant (see below).

Assuming that the life-time of the cool contact configuration is of the
order of one or more Gyr, we consider a stationary solution of the above
set of equations on a time scale longer than the thermal time scale. We
put, then, $\partial/\partial t \equiv 0$. We assume that a steady mass
flow from the hotter primary to the secondary exists in the common
envelope.  The assumption of steady motion eliminates any dynamical
instabilities, generation of acoustic waves or any other processes taking
place on a very short time scale. After leaving the neck between the
components the mass stream quickly assumes the azimuthal motion with
hydrostatic equilibrium condition in the vertical and meridional directions
satisfied.

Let us first consider an inviscid motion. The inertial term on the left
hand side of Eq.~(2) is of the order of $v^2/l = v^2/2\pi R_{\mathrm{sec}}$,
as the flow driving force results from the azimuthal pressure gradient. It
can be compared to the Coriolis term of the order of $2\Omega v$.
Their ratio is called the Rossby number $Ro$

\begin{equation}
Ro = \frac{v^2}{2\pi R_{\mathrm{sec}}}/2\Omega v \approx \frac{v}{4\pi v_{\mathrm{eq}}}\,,
\end{equation}

where $v_{\mathrm{eq}}$ is the equatorial velocity of a rotating star. In
  a typical W UMa type star $v_{\mathrm{eq}} \approx 100-150$ kms$^{-1}$, so as
  long as the flow is not highly supersonic the Rossby number is
  small, i.e. $Ro << 1$, and we can safely
  neglect the inertial term. The equation of motion becomes then

\begin{figure}
\begin{center}
\rotatebox{0}{\scalebox{0.48}{\includegraphics{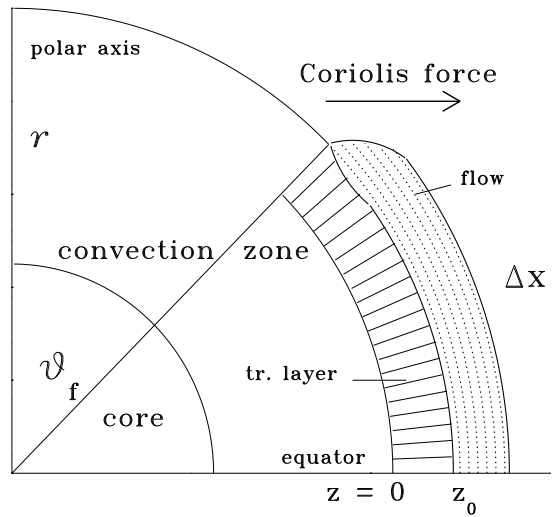}}}
\caption{\label{f2}
The reference frame and geometry of the flow in the meridional plane
  of the secondary component. The stream flows away, 
perpendicularly to the plane
  of the paper, in the $\phi$-direction.  Scales for the transition layer depth and the 
flow thickness have been exaggerated.}
\end{center}
\end{figure}

\begin{equation}
 \frac{1}{\rho}\mathbf{\nabla}p =  2\mathbf{\Omega\times v} +
 \mathbf{\nabla}\varphi\,, 
\end{equation}

where ${\bf v}$ has only one, non-vanishing azimuthal component $v$. The
above equation describes the so called geostrophic flow in which the
meridional component of the Coriolis force balances the lateral pressure
gradient, perpendicular to the direction of motion. Such flows are often
observed in the terrestrial atmosphere when air circulates around a high or
low pressure center. In our case, the Coriolis force confines the mass
  flow to the equatorial region of width depending primarily on the
  flow velocity (see Sect. 3.4 below).

\section{The stream structure}

\subsection{Lateral equilibrium}

We introduce a spherical system of coordinates with the origin at the
secondary star center ($r, \vartheta$, $\phi$) and we neglect the deviation
of equipotential surface shapes from spheres (Fig.~2). Because the stream is
symmetric relative to the equator, we consider only a hemisphere $0 \le
\vartheta \le \pi/2$. Far from the neck matter flows along the
equator of the secondary in a belt with a half-width $\Delta x =
R_{\mathrm{sec}}(\pi/2 - \vartheta_{\mathrm{f}}(r))$, where
$\vartheta_{\mathrm{f}}(r)$ is the depth dependent polar angle of the flow
boundary. In general, pressure inside the flow is a function of all three
coordinates but within our assumptions it is a slowly varying function of
the azimuthal direction so it can be assumed constant when considering
equilibrium in the radial and meridional direction. The equilibrium condition
results from Eq.~(5):

\noindent meridional component

\begin{equation}
\frac{1}{r}\frac{\partial p}{\partial\vartheta} = 2\rho kc_s\Omega
\cos\vartheta\,,
\end{equation}

\noindent and the radial component

\begin{equation}
\frac{\partial p}{\partial r} = 2\rho kc_s\Omega\sin\vartheta
+\rho\frac{\mathrm{d}\varphi}{\mathrm{d}r}\,.
\end{equation} 

Here $c_s$ and $k$ are the sound velocity and the Mach number of the flow,
respectively. We assume $k \le 1$ throughout the paper. Eq.~(6) can be
integrated

\begin{equation}
p(\vartheta) = p_o -2r\rho kc_s\Omega(1 - \sin\vartheta)\,,
\end{equation}

where $p_o$ is the pressure at the equator. The ambient
pressure under the secondary's photosphere is equal to
$p_{\mathrm{sec}}$, so the boundary will be reached when
$p(\vartheta_{\mathrm{f}}) = p_{\mathrm{sec}}$. This gives a condition
for $\vartheta_{\mathrm{f}}$

\begin{equation}
\sin\vartheta_{\mathrm{f}}= 1 - \frac{p_o - p_{\mathrm{sec}}}
{2\rho kc_sv_{\mathrm{eq}}}\,.
\end{equation}

For parameter values characteristic of W UMa stars, typical stream
  widths are around 15$^{\mathrm o}$ - 60$^{\mathrm o}$ (see Sect. 3.4).

Above the photosphere  $p_{\mathrm{sec}} \equiv 0$.
With a total thickness $\Delta r << R_{\mathrm{sec}}$ the stream layer is
thin, so we can replace coordinate $r$ appearing explicitly in Eq.~(8) by
$R_{\mathrm{sec}}$.

If we now include the effects of viscosity, this will produce a drift of
the stream matter into high astrographic latitudes. Molecular viscosity is
negligibly small but the convective viscosity may play a role. We can
estimate a possible drift resulting from that viscosity

\begin{equation}
\nu = \frac{1}{3}l_{\mathrm{conv}}v_{\mathrm{conv}}\,,
\end{equation}

where $l_{\mathrm{conv}}$ and $v_{\mathrm{conv}}$ are the mixing length and
convective velocity near the bottom of the stream.

Assuming that $l_{\mathrm{conv}}$ is of the order of pressure scale height
near the bottom of the flow one obtains from the model of the convective
zone $H_p \approx 10^9$ cm and $v_{\mathrm{conv}} \approx 2\times 10^4$
cms$^{-1}$ so $\nu \approx 7\times 10^{12}$ in cgs units
\citep{bt66,spr74}. The resulting viscous term in the momentum equation is
of the order of $\nu v/\Delta x^2$. For $v \approx 0.3c_s$ and $\Delta x
\approx R_{\mathrm{sec}}/2$ this term is equal to $2\times 10^{-2}$
cms$^{-2}$. It takes about 3 d for a given flow element to encircle the
secondary assuming a typical flow velocity about 10 times slower than
$v_{\mathrm{eq}}$, see below. The total drift distance of the stream in the
meridional direction is equal then to $\sim 2\times 10^3$ km which is
negligible compared to the stream width. We conclude that
viscosity can be neglected altogether when considering dynamics of the mass
flow around the secondary.

\subsection{Vertical equilibrium}

A typical W UMa type binary consists of two unequal mass components
possessing convective envelopes. Except for stars with spectral type around
F0 where the convection zone is very thin and mostly super-adiabatic, its
dominant part is 
stratified nearly adiabatically i. e. according to the pressure-temperature
relation

\begin{equation}
p=KT^{\alpha}\,,
\end{equation}

where $T$ is temperature, $\alpha$ = 2.5 for fully ionized gas and $K$ =
const. throughout the considered part of the convection zone. The parameter
$K$ is called the adiabatic constant and its value depends on the specific
entropy of the matter in the convective zone. Realistic models of the
convective envelope take into account all important effects influencing its
stratification, like variability with depth of $K, \alpha$, and specific
heats as well as deviations from strict adiabaticity \citep{bt66,
spr74}. However, most of these effects take place in the uppermost layers
of the convection zone where the pressure is low. Because we are interested
in the pressure equilibrium at some depth below the stellar surface, a
detailed structure of these shallow layers is unimportant. As the accurate
models of the convection zone of solar type stars indicate, the
stratification becomes very close to adiabatic already for temperatures
higher than about $10^4$ K. We adopt the adiabatic stratification for the
convective envelopes of both components with different values of the
adiabatic constant.

\begin{figure}
\begin{center}
\rotatebox{0}{\scalebox{0.48}{\includegraphics{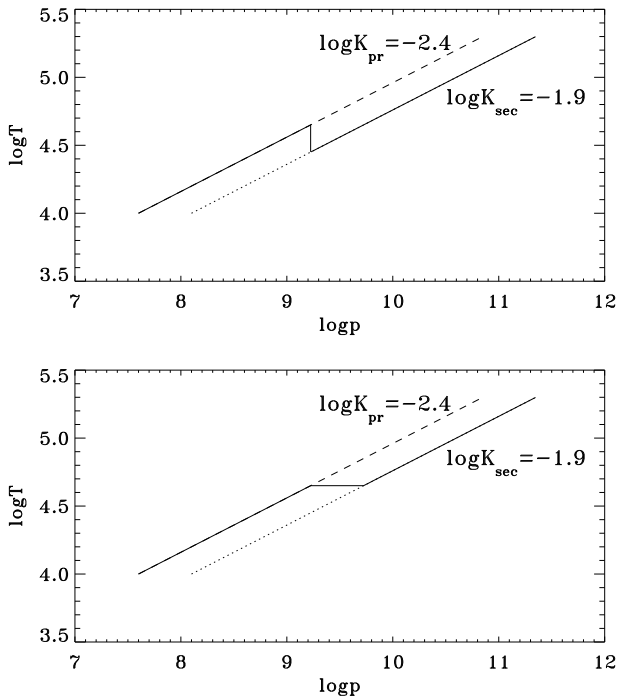}}}
\caption{\label{f3}
The adiabatic $p-T$ relations for the stream matter (upper lines) and the
secondary's convection zone (lower lines). The condition of the 
continuous pressure across the contact
surface between the stream and the unaffected convective layer beneath 
requires a temperature
discontinuity (a solid line, top). The formation of a radiative 
transition layer with a flat
temperature gradient between the stream and the deeper unaffected 
layers results in a continuous transition
both in pressure and temperature (a solid line, bottom).}
\end{center}
\end{figure}

Matter flowing from the primary is characterized by the same adiabatic
constant $K_{\mathrm{pr}}$ as the convection layer of the primary. The
convection layer of the secondary is described by $K_{\mathrm{sec}}$.  In
general, $K_{\mathrm{sec}} < K_{\mathrm{pr}}$. Figure 3 gives an example of
two pressure-temperature relations corresponding to two adiabatic constants
differing by 0.5dex. The upper line on each diagram gives the $p-T$ relation
for the stream and the lower line for the ambient convection layer of the
secondary.

Vertical hydrostatic equilibrium condition requires continuity of the
pressure at the boundary between the stream and the underlying convective
layer not affected by the flow. Because the secondary convection zone has
specific entropy lower than the stream, a discontinuity in entropy and
temperature must occur at this boundary (a vertical segment of the solid
line in Fig.~3 top). A steady state model with a contact discontinuity of
this kind was considered by \citet{shu76, shu79}. They assumed that matter
streaming from the primary fully covers the secondary component of a
contact binary. They tried to build a model in which the nuclear energy of
the secondary can flow outwards in spite of the
discontinuity. \citet{haz78} argued, however, that such a model contradicts
the second law of thermodynamics because the nuclear energy of the
secondary cannot flow from the low to the high entropy medium
above. Instead, the nuclear energy will heat the convection zone on a
thermal time scale rising its specific entropy to the value characteristic
of the blanketing matter, just as assumed in the TRO model.

The situation considered in the present paper is, however, entirely
different. The streaming matter covers now only a part of the
secondary. There is no need to introduce a contact discontinuity. The hot
stream matter heats the layer lying underneath and lowers the temperature
gradient there. As a result, the convective energy flow from below is
inhibited. In steady state a radiative layer is formed with a low
temperature gradient and specific entropy increasing outwards from the
value characteristic of the secondary convection zone up to the value
characteristic of the stream gas. We will call it a transition layer. Such a
layer with the temperature gradient equal to zero is shown in Fig.~3
(bottom) as a horizontal segment of the solid line. In real stars the
gradient will probably be not equal exactly to zero so the transition layer
will be thicker than shown. The heat transport is completely, or nearly
completely, blocked in the transition layer which acts as an insulator. The
energy blocked by the stream will be redistributed over the whole
convection zone of the secondary and ultimately radiated away from the
polar regions not covered by the stream matter 
(see Section 4).

To consider the vertical structure of the stream in more detail we need to
calculate $p_o - p_{\mathrm{sec}}$ with depth. Because the depth of the
layers in question is much lower than the stellar radius we can use the
plane parallel approximation with the coordinate $z$ replacing $r$
  (Fig.~2). Let us
assume that the secondary convection zone remains unaffected by the stream
below the bottom of the transition layer and we put the reference level $z
= 0$ there .  The pressure is constant over the whole equipotential surface
lying at that level.  We denote it by $p_b$.  Within the isothermal
transition layer pressure varies as

\begin{equation}
p(z)=p_b\exp(-z/H_p),\quad H_p =
\frac{k_{\mathrm{B}}T_o}{\mu\mathrm{H}g_{\mathrm{eff}}},\quad 0\le z\le z_o\,, 
\end{equation}

where $\mu$ is the mean molecular weight, H -- mass of the hydrogen atom,
$k_{\mathrm{B}}$ -- the Bolzmann constant, $T_o$ -- temperature of the
layer, $g_{\mathrm{eff}}$ -- the effective gravity and $z_o$ corresponds to
the bottom of the stream, which agrees with the top of the transition
layer. The pressure is continuous at the interface between the transition
layer and the adiabatically stratified stream where temperature varies as

\begin{equation}
T(z) = T_o - \frac{g_{\mathrm{eff}}}{c_p}(z - z_o),\quad z\ge z_o\,,
\end{equation}

where $c_p$ is specific heat at constant pressure. 
When calculating the $T(z)$ relation we neglect
variability of the effective gravity. With $T(z)$
known, the vertical pressure stratification is obtained from Eq.~(11) where
$K = K_{\mathrm{str}}$.  As it will be shown below, the entropy of the
stream matter varies very little during the flow around the secondary, so
we can put $ K_{\mathrm{str}} \approx K_{\mathrm{pr}}$ everywhere.

\begin{figure}
\begin{center}
\rotatebox{0}{\scalebox{0.48}{\includegraphics{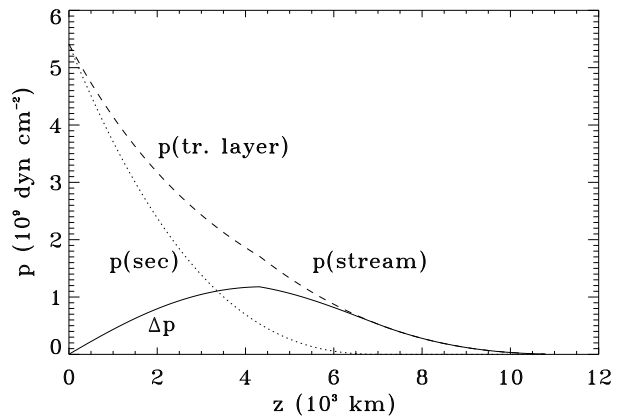}}}
\caption{\label{f4}
The vertical pressure distribution across the
stream together with the transition layer (broken line), in the 
unperturbed secondary
convection zone (dotted line) and their difference (solid line),
balanced by the meridional component of the Coriolis force (see text). The
bottom of the transition layer is at $z = 0$ and the bottom of the stream
(expected to coincide with the Roche critical surface) is at $z
\approx 4\times 10^3$ km.} 
\end{center}
\end{figure} 

It is reasonable to assume that in steady state the transition layer will
be partially dragged by the stream in such a manner that a gas layer
lying immediately below the stream moves with  a stream velocity and 
deeper layers move slower until zero velocity is attained at the bottom
of the transition layer. We arbitrarily assume that velocity inside the
transition layer varies as

\begin{equation}
v(z) = kc_s\frac{\rho_b - \rho(z)}{\rho_b - \rho_t}\,,
\end{equation}

where $\rho_b$ and $\rho_t$ are the densities at the bottom and at the top
of the transition layer. Thus defined velocity vanishes at the bottom and
reaches the stream velocity at the top.

Far from the stream the convection zone of the secondary remains unperturbed
with the
adiabatic stratification.  Neglecting a difference between $c_p$ in
the stream and in the ambient convection zone we have

\begin{equation}
T(z) = T_o - \frac{g_{\mathrm{eff}}}{c_p}z\,,
\end{equation}

%where $g_{\mathrm{sec}}$ is the surface averaged gravity on the secondary
%component. It differs from $g_{\mathrm{eff}}$ mainly because of the
%non-zero vertical component of the Coriolis force (see Eq.~(7)). In most
%cases this component is small compared to gradient of gravitational
%potential so we can use the same quantity in Eqs.~(13) and (15).  
Pressure
stratification is obtained again from Eq.~(11), except that now
$K = K_{\mathrm{sec}}$.

Figure~4 shows an example of the vertical pressure distribution inside the
transition layer and the
stream with depth equal to 1 \% of the primary's radius (broken line), in the
unperturbed secondary convection zone (dotted line) and their difference
(solid line). As it is seen from the figure, the stream
extends above the surface of the secondary and produces an equatorial
bulge held by the meridional component of the Coriolis force (see also
Fig.~5).  

\subsection{The thermal stream structure along the flow}

With the depth and width of the stream specified, the mass transfer rate
$F_{\rho}$ can be calculated

\begin{equation}
F_{\rho} = 2\Delta x\int_{0}^{\Delta r}kc_s\rho{\mathrm d}r = 2\Delta
xkc_s\overline{M}_{\Delta r}\,,
\end{equation}

where $\overline{M}_{\Delta r}$ is the column mass above the level $\Delta
r$. The stream also carries thermal energy which is partly radiated away
during its flow around the secondary. We will now discuss the
variation of the stream thermal structure along the flow.

Let us consider a slice of matter perpendicular to the flow with thickness
$R_{\mathrm{sec}}{\mathrm{d}}\phi$. Its total heat (internal energy plus
enthalpy) capacity d$Q$ is equal
to

\begin{equation}
{\mathrm d}Q=2\Delta xR_{\mathrm{sec}}{\mathrm d}\phi\int_{0}^{\Delta r}
\rho c_pT{\mathrm d}r = 2\Delta xR_{\mathrm{sec}}c_p\overline{MT}_{\Delta
  r}{\mathrm d}\phi\,, 
\end{equation}

where

\begin{equation}
\overline{MT}_{\Delta r} = \int_{0}^{\Delta r}\rho T{\mathrm d}r\,.
\end{equation}

The matter flowing around the secondary radiates energy at a rate $\sigma
T_{\mathrm{str}}^4$ per unit time and area, where $T_{\mathrm{str}}$ is the
effective temperature of the stream. In general, $T_{\mathrm{str}}$ will
decrease along the flow from its initial value, assumed to be close to the
effective temperature of the primary, $T_{\mathrm{pr}}$, and its precise
variation could be determined by solving the transfer equation within the
stream. To avoid cumbersome calculations we discuss two limiting cases:
first, when the total heat capacity of the stream is much higher than the
total energy radiated away, and then, when the heat capacity is
comparable to the radiated energy.

The slice has a radiating area $2\Delta xR_{\mathrm{sec}}{\mathrm{d}}\phi$
and emits d$F_{\mathrm{rad}} = 2\sigma T_{\mathrm{str}}^4\Delta
xR_{\mathrm{sec}}{\mathrm{d}}\phi$ energy per unit time. It takes a time
equal to $2\pi R_{\mathrm{sec}}/kc_s$ to encircle the secondary. The
total energy radiated away during this time is

\begin{equation}
{\mathrm{d}}L = \frac{4\pi\sigma T_{\mathrm{str}}^4\Delta
  xR_{\mathrm{sec}}^2{\mathrm{d}}\phi}{kc_s}\,. 
\end{equation}

From Eqs.~(17) and (19), the limiting case of large heat capacity,
$\mathrm{d}Q >> \mathrm{d}L$,
corresponds to the condition

\begin{equation}
c_p\overline{MT}_{\Delta r} >> 
\frac{2\pi\sigma T_{\mathrm{pr}}^4R_{\mathrm{sec}}}{kc_s}\,.
\end{equation}

If the radiated energy is small compared to the total heat capacity of the
stream, its effective temperature will change very little from its initial
value $T_{\mathrm{pr}}$ so we can put $T_{\mathrm{str}} \approx
T_{\mathrm{pr}}$. In case of low heat capacity the effective temperature of the
stream will approach the effective temperature of the secondary
and the stream matter will mix together with its
convection zone.

Assuming that the energy outflow from the stream influences uniformly its
entropy, i. e. that the stream specific entropy is constant with depth, we
can calculate its decrease from Eq.~(3). In case of the large heat capacity
we obtain

\begin{equation}
\Delta s = - \frac{4\pi R_{\mathrm{sec}}\sigma
  T_{\mathrm{pr}}^4}{kc_s\overline{MT}_{\Delta r}}\,.
\end{equation}

In case of the low heat content, $T_{\mathrm{pr}}$ in Eq.~(21) should be
replaced by the properly defined average stream effective temperature.

The entropy decrease can be compared with entropy of the primary
convection zone

\begin{equation}
s_{\mathrm{pr}} = - \frac{k}{\mu\mathrm{H}}\ln K_{\mathrm{pr}}\,.
\end{equation}

After return to the primary, the stream matter sinks in its convection
zone.

\begin{figure}
\begin{center}
\rotatebox{0}{\scalebox{0.48}{\includegraphics{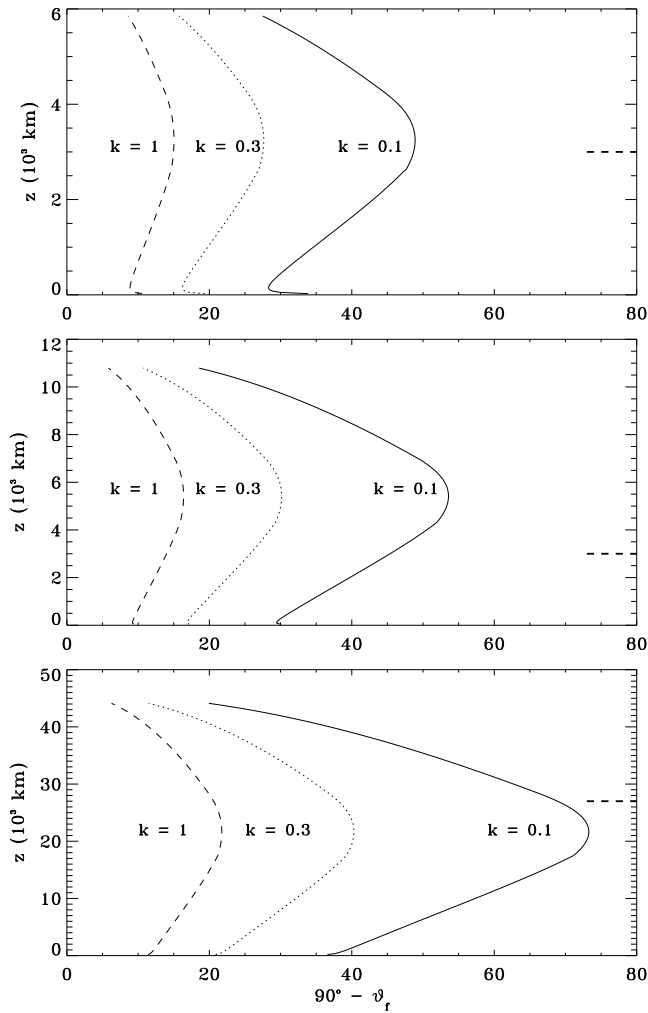}}}
\caption{\label{f5}
The width of the stream as a function of height above the bottom of the
transition layer for three values of $k$ describing the ratio of stream
to sound velocity and for three depths of the stream: 0.5 \% (top), 1
\% (middle) and 3 \% (bottom) of the primary's radius. A broken heavy line on
the right indicates an approximate level of the secondary's photosphere. Note
a weak dependence of the stream width on its total depth.}
\end{center}
\end{figure} 

\subsection{Numerical example: AB And}

To obtain a numerical estimate of the parameters describing an actual
stream in a cool contact binary we apply the above derived equations to a
typical W-type contact binary. As an example, we take AB And. It has the
orbital period $P_{\mathrm{orb}} = 0.33$ d, component masses
$M_{\mathrm{pr}} = 1.04 M_{\odot}$ and $M_{\mathrm{sec}} = 0.60 M_{\odot}$,
and volume radii $R_{\mathrm{pr}} = 1.02 R_{\odot}$ and $R_{\mathrm{sec}} =
0.78 R_{\odot}$ \citep{bar04}. These data give $v_{\mathrm{eq}}$ = 120
kms$^{-1}$ for the secondary component. The star has a relatively low
fill-out factor $f = 0.05$ \citep{bar04}, which corresponds to about 1 \% of
the primary star radius. Assuming that the equal pressure equipotential
surface is close to the inner critical Roche lobe we obtain $\Delta r
\approx 0.01$ for the depth of the stream. The sound velocity is equal to
30 kms$^{-1}$ at this depth. The width of the stream can be calculated from
Eq.~(9). Figure 5 shows the results. Here the angular half-width is plotted
as an abscissa for different heights above the bottom of the transition
layer. The broken heavy line on the right indicates the approximate
photospheric level of the secondary. The diagrams give the stream width for
three values of $\Delta r/R_{\mathrm{pr}}$ equal to 0.005 (top), 0.01
(middle) and 0.03 (bottom) and three values of $k$, as indicated. The width
was calculated up to the level where temperature reached 10 000 K because
for lower temperatures the adiabatic model breaks down.  As we see, the
largest width of the stream is close to the photosphere of the secondary
but the Coriolis force produces a bulge above the stellar equator with a
height of the order of the stream depth. As expected, a lower stream
velocity results in its increased width and in the limit of zero velocity
the stream would cover the whole secondary. For flow velocities comparable
to the sound velocity, i. e. for $k$ between 1 and 0.3, the stream has a
half-width between 15$^{\mathrm o}$ and 30$^{\mathrm o}$ (Fig.~3), and
covers between 1/4 and 1/2 of the stellar surface, respectively.

A schematic side view of AB And with a stream covering 50 \% of the
secondary's surface is shown in Fig.~6.

\begin{figure}
\begin{center}
\rotatebox{0}{\scalebox{0.48}{\includegraphics{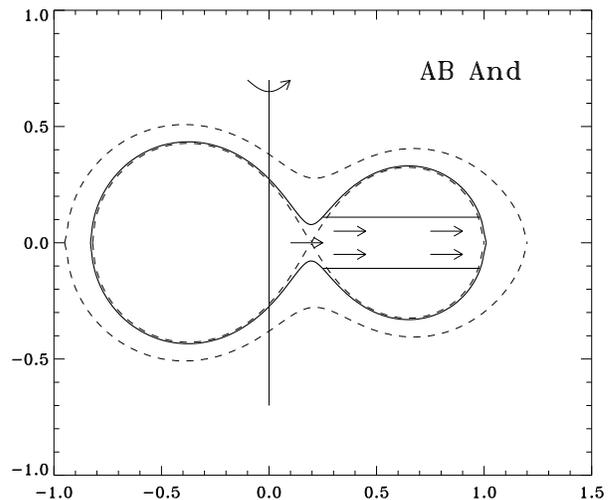}}}
\caption{\label{f6}
A side view of AB And with a stream covering a half of its surface. Broken
lines indicate inner and outer critical surfaces and the solid line shows
the binary surface.}
\end{center}
\end{figure}

The stream with a depth of $0.01$ carries enough mass and energy to be in
the regime of large heat capacity. To see it we substitute the following
values to Eqs.~(17) and (20): $\Delta r = 7\times 10^{10}$ cm, $\overline
c_p = 10^9$ (in cgs units), $T_{\mathrm{pr}}$ = 5590 K (see next section)
and $k$ = 0.3. The column mass and the integral $\overline{MT}_{\Delta r}$
are calculated from the convection zone model. We obtain
$\overline{M}_{\Delta r} = 3.5\times 10^5$ gcm$^{-2}$ and
$\overline{MT}_{\Delta r} = 1.7\times 10^{10}$ gKcm$^{-2}$. The resulting
mass transfer rate is $F_{\rho} = 5\times 10^{-4} M_{\odot}$/year. The left
hand side of Eq.~(19) gives $1.7\times 10^{19}$ (in cgs units) and the
right hand side gives $4\times 10^{16}$ in the same units. The resulting
ratio of both numbers is equal to $2.4\times 10^{-3}$ which indicates that
the non-equality given by Eq.~(19) is very well fulfilled. The numbers
become comparable only for stream depths less than 0.5 \% of the stellar
radius. Figure~5 (top) shows the calculated width of the stream with the
depth of 0.5 \%. It is somewhat surprising that its width is not much
different from the widths of much more massive streams shown in Fig.~5
(middle) and (bottom). It is clear that only extremely marginally contact
binaries with overfill factors less than a couple of hundredths will have
streams with surface temperature approaching the temperature of the
uncovered part of the secondary component.

The change of specific entropy of the stream in AB And can be calculated
from Eq.~(20). With the adopted parameters we obtain $\Delta s = -
1.2\times 10^6$ ergg$^{-1}$K$^{-1}$. According to Eq.~(21) the initial
value of the
specific entropy is equal to $4.59\times 10^8$ in the same units so the
decrease of entropy during the flow around the secondary is small
indeed. It is possible then that the stream returning to the primary
component will not sink immediately after passing the neck but it may
move over a substantial fraction of the primary's circumference before it
plunges in the convection zone.

To summarize, we see that the Coriolis force confines the near sonic stream
to the equatorial belt of width $\pm (15^{\mathrm{o}}-30^{\mathrm{o}})$
around the secondary. The stream covers 25-50 \% of the stellar
surface. For a primary overflowing its Roche lobe by $\sim 1 \%$ or more
the stream radiates a very small fraction of its total heat capacity which
results in its nearly constant surface temperature equal to the primary's
temperature.  Substantially shallower streams, possible in contact binaries
with an extremely marginal contact, will show a temperature decrease from
the initial value close to the primary's temperature on the trailing
hemisphere of the secondary, down to the value close to the secondary's
temperature on the leading hemisphere of the star.

After return to the primary, the stream will sink in its convection zone
due to the lower entropy. The Coriolis force deflects the returning stream
to the opposite side of the neck, compared to the stream flowing towards
the secondary. The deflection should reduce an interaction of both flows
although it will not separate them completely.  The returning flow will
collide with matter streaming towards the neck (see Figs.~1 and 2 in
\citealt{oka02}). However, the collision will not be head-on, as obtained
by \citet{mar95} who simulated a two dimensional motion in an equatorial
plane of a contact binary. Geometry applied by the authors prevents matter
from crossing the plane containing centers of both stars and the rotation
axis, so the stream cannot move to one side of the neck, as shown by
three dimensional calculations \citep{oka02}.

\section{Global reaction of the binary to the mass and energy flow}

\subsection{The influence of the stream}

\begin{figure}
\begin{center}
\rotatebox{0}{\scalebox{0.48}{\includegraphics{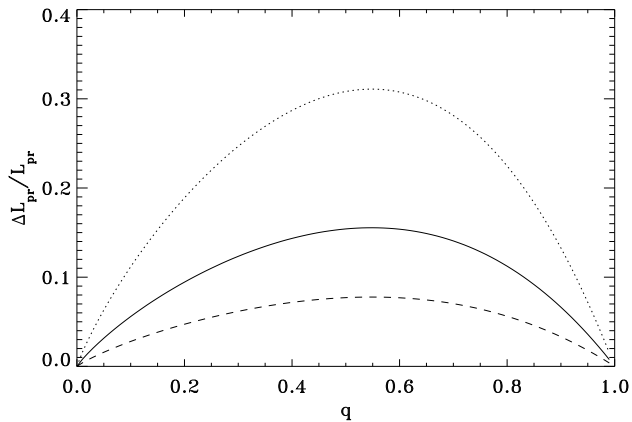}}}
\caption{\label{f7}
A fraction of the primary component luminosity radiated away by the stream
flowing around the
secondary component as a function of mass ratio. The curves correspond to 
three values of coverage of the secondary by the stream: 
full coverage (dotted line), half-coverage (solid line) and quarter-coverage
(broken line).} 
\end{center}
\end{figure}

As it was mentioned earlier, the stream acts like a hot blanket covering
part of the secondary and blocking the energy flow from below by rising the
temperature of the layer immediately beneath. As a result, a transition
radiative layer is formed with a decreased temperature gradient in which
the specific entropy increases from the bottom value characteristic of the
secondary convection zone to the top value characteristic of the
stream. 

What is the global reaction of a star when a fraction of its surface cannot
radiate away the stellar core energy due to an obstacle blocking the energy
outflow? The physical nature of the obstacle is not so important as long as
it occurs in the outermost stellar layers.  One example of such an obstacle
partly blocking the energy flow from below is a dark starspot appearing on
the surface of an active star. Another example can be seen in close binaries
when a hot companion irradiates a part of the surface of a cool component. 
Both effects have
been discussed at length in the literature (e. g. \citealt{rit00},
\citealt{spr86} and references therein).  The blocked energy flux is
redistributed inside the convection zone and re-radiated by the unperturbed
part of the stellar surface. The detailed models show that the specific
entropy of the convection zone is somewhat increased and so is the stellar
radius. For stars with deep convection zones, or fully convective, the
nuclear energy production may also be affected. Assuming a steady state in
which a constant fraction of the stellar surface is covered by spots over a
time long compared to the thermal time scale of the convection zone,
\citet{spr86} obtained the following expression for the equilibrium stellar
radius

\begin{equation}
R_e(f_{ps}) = R_o(1-f_{ps})^d\,,
\end{equation}

where $R_e$ and $R_o$ denote the equilibrium radius of the spotted and
unspotted star, $f_{ps}$ is the fraction of the stellar surface permanently
covered by spots and $d \approx -0.1$. An essentially identical expression
was found by \citet{rit00} for stars irradiated by hot companions. For
$f_{ps}$ = 0.25 and 0.5 the equation gives $R_e = 1.03R_o$ and $1.07R_o$,
respectively.  In stars with not too deep
convection zones the nuclear energy production remains unaffected. The
reduction of the effective radiating surface is exactly compensated by an
increase in surface temperature of the undisturbed part of the star

\begin{equation}
T^e_{\mathrm{eff}} = (1-f_{ps})^{-\frac{1+d}{4}}T_{\mathrm{eff}}\,,
\end{equation}

where $T^e_{\mathrm{eff}}$ and $T_{\mathrm{eff}}$ are the equilibrium and
undisturbed effective temperature.  More accurate values of the equilibrium
radius and temperature depend on stellar mass \citep{spr86}.

The primary component suffers from additional cooling of its surface by the
stream encircling the secondary. The stream feeds the primary with the
lower entropy matter which is mixed with the rest of its convection
layer. As a result, the specific entropy of the convection zone is somewhat
lowered and so is the stellar radius but the expected changes should be
close to those occurring in the secondary component, i.e. they should 
not exceed one,
or at most a few percent. The primary, slightly undersized for its mass, will
appear closer to ZAMS in the HR diagram. Neglecting its radius change we
can obtain the modified effective temperature of the primary

\begin{equation}
T^m_{\mathrm{pr}} = (1-\frac{\Delta L_{\mathrm{pr}}}{L_{\mathrm{pr}}})\,,
\end{equation}

where $L_{\mathrm{pr}}$ is the luminosity of the primary component and
$\Delta L_{\mathrm{pr}}$ is a fraction of this luminosity radiated away by
the stream flowing around the secondary component (Fig.~7).

Now let us calculate the expected values of the thermal parameters of AB
And.  The effective temperature of a single one solar mass star is about
5800 K. That should be close to the temperature of the undisturbed
primary. Assuming that secondaries of W UMa type stars are evolved stars
with hydrogen depleted in their stellar cores \citep{ste04, ste06a} we
obtain the following values of the global parameters for the unperturbed secondary of
AB And: $T_{\mathrm{eff}}$ = 4570 K, and $L/L_{\odot}$ =
0.24. The data were taken from the evolutionary models of stars with masses
equal to the present masses of the secondaries.
This is not fully consistent with the present considerations because,
according to the accepted model, the secondaries were originally more
massive and their present masses result from mass exchange. It is hoped,
however, that differences between the adopted and fully consistent models
are of secondary importance. Let us assume, as an example, that the stream
covers 50\% of the secondary and that its nuclear energy production rate is
unaffected by this. According to Eq.~(23) the steady state effective
temperature of the surface fraction not covered by the stream rises
to 5340 K.  The primary transfers about 16~\% of its luminosity (see
Fig.~7) to the secondary so its actual effective temperature, resulting from
Eq.~(25), drops to 5590 K. The effective temperature of the stream is
of course the same. The ``final'' surface averaged temperature of the
secondary is equal to about 5480 K, which is nearly identical as obtained
by \citet{bar04} from observations who give
$T_{\mathrm{sec}}$ = 5500 K. Such a small difference between the
  predicted and observed temperature is very likely fortuitous, taking into
  account a large observational uncertainty of the observed value (see
  below). However, the observed primary's temperature, 
equal to 5140 K, is {\em substantially} lower than resulting from the present model. 
A possible explanation of this discrepancy is discussed in the next Section.

\subsection{The W-type phenomenon and photospheric starspots}

The observed temperatures of components of a W UMa type binary belong to
the least accurately determined stellar parameters \citep{ye05}. The
primary's temperature is usually determined from the spectral type which is
uncertain due to very broad spectral lines, or from the photometric
index. It is kept constant during the light curve modeling. The
secondary's temperature results from modeling of the light curve, in
particular from the relative depth of both minima, which gives a ratio of
both temperatures. Such an analysis gives only a rough estimate of the
surface averaged temperatures and is insensitive to their possible
variations across the stellar surface. Regarding the relative temperatures
of the components, cool contact binaries are divided into two types: W-type
when secondaries look hotter than primaries and A-type when the opposite
takes place \citep{bin70}. A typical temperature difference between the
components of genuine contact binaries is of the order of a few hundred
kelvin \citep{ye05}. A few stars alternating between W-type and A-type are
known with a temperature difference always staying close to zero. The
present model gives $\Delta T = T_{\mathrm{pr}} - T_{\mathrm{sec}}$ equal
to 110 K for AB And. Several A-type stars are known with
$\Delta T$ close to this value. However, AB And is of W-type and
this phenomenon cannot
be explained within the energy transfer mechanism alone. No such mechanism,
obeying basic laws of physics, can transfer energy from cooler to hotter
medium in an isolated system. So, the most popular explanation of the
W-type phenomenon assumes that dark, cool spots cover a substantial
fraction of the primary's surface \citep{ruc93}. The temperature of its spot-free part may
be higher than the secondary's temperature but the apparent surface
averaged temperature is lower. W UMa-type stars have a very high level of
chromospheric-coronal activity \citep{cd84, ruc85, vil87, ste01} and
variations of the light curve, observed on time scales of decades, indicate
that they also possess dark starspots covering a variable fraction of the
stellar surface \citep{gz06, cou08}. Perhaps the most extreme example of light curve
variability due to a variable coverage by spots can be seen in
\citet{rp02}.

We can estimate how large should be the fraction of the primary's surface
covered by dark spots to lower its average temperature to a required
value. Assuming, for simplicity, that umbrae have surface brightness of
1/4 of the undisturbed photosphere (it is so in the solar case) and that
they dominate the spots we have

\begin{equation}
T_{\mathrm{pr}}^{\mathrm{f}} = (1-0.75f_s)^{1/4}T_{\mathrm{pr}}\,, 
\end{equation}

where $T_{\mathrm{pr}}^{\mathrm{f}}$ is the ``final'' primary's temperature
modified by spots and $f_s$ is the spot coverage. It is assumed here that
spots appear and disappear on a time scale short compared to the thermal
time scale of the convection zone. When the spots appear, they simply block
the respective energy flux which does not reappear elsewhere. In other
words, we consider now only the variable component of the spot coverage. If
there also exists a permanent spot component covering a fraction $f_{ps}$
of the primary's surface it will influence the effective temperature as
discussed in Sect. 4.1.

For $f_s$ = 0.25 we obtain $T_{\mathrm{pr}}^{\mathrm{f}}$ = 5310 K, hence
$\Delta T$ = -170 K. When $f_s$ is increased to 0.5, we obtain
$T_{\mathrm{pr}}^{\mathrm{f}}$ = 4970 K, hence $\Delta T$ = -510 K. This
should be compared with the observed value $\Delta T$ = -340 K.  As we
see, the required spot coverage is quite high but reasonable. There are a
number of single, rapidly rotating stars with a variable spot coverage
within this range, e. g. AB Dor \citep{inn08} or BO Mic
\citep{wol08}. Heavily spotted are also cool detached binaries with periods
shorter than one day, like XY UMa \citep{pri01} or BI Cet \citep{cut03}.

The absolute magnitude of AB And, obtained from the visual magnitude and
the distance is equal to 4.05 mag \citep{rd97}.  With BC = -0.08 mag
\citep{all73} the binary has the luminosity $L \approx
1.6L_{\odot}$. Adopting $L \approx 1.16L_{\odot}$ for the MS primary of AB And
and $L \approx 0.25L_{\odot}$ (see above) for the evolved secondary we
obtain $1.4L_{\odot}$ for the total luminosity of the binary. This is less
than observed but the difference is not large and can be due to
observational uncertainties.  Note that if the secondary were a normal MS
star, its core luminosity would be close to $0.1L_{\odot}$ which increases the
discrepancy with observations.

\subsection{The case of AW UMa}

Up to recently the spectroscopic observations of contact binaries were not
accurate enough to resolve an additional flow on a rapidly rotating
component. High-precision observations obtained in the last years by
S. M. Rucinski and his group changed the situation. \citet{pruc08} analyzed
very accurate profiles of spectral lines of AW UMa -- an A-type contact
binary with an extreme mass ratio. The observations indicate the existence
of an equatorial bulge on both components, moving with velocity 20-30 km/s
relative to the rotating frame of reference. The present model predicts
such a bulge on the secondary star. Although dynamics of the returning
stream was not considered in the present paper it was argued
that the returning flow may travel a
significant fraction of the circumference of the primary before it
disappears beneath the stellar surface. The returning flow, together with
the stream formed on the leading hemisphere of the primary, will produce a
spectral feature similar to signature of an equatorial bulge.

The observations by \citet{pruc08} showed in addition that parts of the
line profiles formed away from the equatorial regions of AW UMa are too
narrow for the rigidly rotating components filling the Roche lobes.
The authors suggest that the components may be undersized by about 15
percent. However, another explanation is also possible. The flow pattern
obtained by \citet{oka02} from hydrodynamic simulations of a semi-detached
binary shows the presence of giant eddies surrounding the poles of the
donor and rotating contrary to the orbital motion (i. e. contrary to the
direction of the stream). Such eddies {\em reduce} rotational broadening of
lines formed in the polar regions. In a contact binary similar eddies may
also occur on the secondary. If so, the observed profiles mimic rigid
rotation of undersized stars. 

\citet{pruc08} stress, however, that such
profiles are not common and are not observed e.~g. in another contact
binary, V556 Oph, in which the spectroscopic observations are in agreement
with a conventional contact model. The difference between line profiles
observed in AW UMa and V556 Oph may result from a different
evolutionary status of both stars and different size of the stream. The
secondary component of AW UMa is very likely to be a highly evolved star with
a massive helium core and a very tenuous envelope \citep{pacz07}. The
evolutionary computations suggest that the size of the secondary may be
close to its maximum, or even past it, so the secondary may presently
undergo evolutionary shrinking. In such a case, the stream can be more
massive and move faster than in other binaries in which secondaries still
expand due to evolutionary effects. A contact configuration of AW UMa may
still be sustained by a slow (net) mass transfer from the primary to the
secondary. This transfer results in a shortening of orbital
period i. e. it acts in the same direction as AML.  The situation when two
operating mechanisms tighten the orbit cannot last for long. If so, AW UMa
is in an exceptional evolutionary state just prior to final merging of both
components. It would be very interesting to identify other stars with
properties similar to AW UMa. Secondaries with little or no helium core
expand as they evolve. A self-regulating mechanism works in such binaries
\citep{gs08}. The expanding secondary transfers mass to the primary on an
evolutionary time scale. Mass transfer from a less massive to a more
massive component widens the orbit. At the same time, AML tightens the
orbit on approximately the same time scale. The whole process contains a
negative feed-back: too fast secondary expansion results in an increased
mass transfer, which leads in turn to an appropriate widening of the orbit
(in spite of AML) and the Roche lobe increase. The Roche lobe of the
secondary approaches its surface and the mass transfer is cut down.  If,
for some reason, the expansion rate of the secondary becomes low, the
orbit tightens due to AML, the Roche lobe shrinks and the higher mass
transfer rate is restored. Similarly, if AML significantly increases
(decreases), the Roche lobe varies correspondingly so that mass transfer
rate increases (decreases), resulting in an orbit which is
almost insensitive to such fluctuations. V556 Oph and many other W UMa type
stars are very likely in such an evolutionary state. Their streams will be
less violent and not so contrasted as in case of AW UMa.

\section{Summary and discussion}

\subsection{The essentials of the circulation model}

It was shown by \citet{ste04, ste06a, ste06b} that an evolutionary
model of a cool contact binary, in which the primary component is a MS star
and the secondary component is in an advanced evolutionary stage with a
hydrogen depleted core, fulfills simultaneously two conditions: both
components are in thermal equilibrium and their sizes conform to the
geometrical requirement of the Roche model. Energy exchange between the
components was not a part of that model. The present paper considers the
problem of the energy transport. 

A common envelope of a contact binary with unequal components cannot achieve
hydrostatic equilibrium. Unequal heating from below of the base of the
common envelope by the emergent energy fluxes of the components produces
temperature difference on each equipotential surface. A resulting
baroclinic structure in the common envelope drives large scale mass motions
between the components on a dynamical time scale. Because the stream
transports thermal energy together with mass, stellar luminosities are
redistributed over the common surface. 

Three dimensional hydrodynamic simulations of the mass loss in a
semi-detached binary demonstrate that a stream leaving the donor star is
deflected by the Coriolis force in the direction of the orbital motion so
that matter flows at an angle of about 10$^{\mathrm o}$ to the line joining
star centers \citep{oka02}. The speed of the stream reaches sound speed already in the
vicinity of L$_1$ point \citep{ls75, oka02}. After leaving the L$_1$ region
the stream moves down the potential well. If the companion is compact
enough, the stream misses its surface, encircles it and forms a disc. If a
radius of the companion is larger than a certain critical value, the stream hits
its surface. The subsequent stream motion is restricted to two dimensions
along the stellar surface. A contact configuration is an extreme case of a
large companion filling the same equipotential surface as the donor. The
stream is then forced to move from the beginning along the common
equipotential surface, to encircle the other component and to return to
donor. The same Coriolis force which deflects the stream, prevents it now
from spreading up to the stellar poles. Even in the presence of turbulent
viscosity the spreading is insignificant. The stream flows in an equatorial
belt with the width determined by the stream velocity. Its part is raised
above the photosphere and forms an equatorial bulge with height comparable
to the stream depth.

Assuming that the bottom of the stream is close to the inner critical Roche
surface we can calculate the global parameters of the flow. Typical
overfill factors of the W UMa type binaries correspond to thickness of the
common envelope of the order of one or a few percent of the stellar radius
\citep{ye05}. For the stream depth of one percent and velocity close to
the sound speed the stream has a total width of 30$^{\mathrm o}$-60$^{\mathrm
o}$ and carries about $5\times 10^{-4} M_{\odot}$/year with velocity of
10-30 km/s. Its heat capacity is a few orders of magnitude higher than the
energy radiated away during the flow around the secondary. As a result, the
effective temperature of the stream and its specific entropy decrease very
little when it returns to the primary. The returning flow may travel a
significant fraction of the stellar circumference before it sinks. Only
binaries with an extremely marginal contact in which the common envelope is
thinner than about 0.5 percent of the stellar radius have streams with heat
capacity comparable to the energy radiated away. The effective temperature
and specific entropy of such a stream approach the ambient temperature and
entropy of the secondary's convection zone.

The stream covering a fraction of the secondary modifies its thermal
structure. It heats the matter immediately beneath reducing the temperature
gradient. The reduced gradient effectively blocks the energy flow from
below. A similar situation occurs in stars covered by dark spots or
irradiated by hot companions. The blocked energy is redistributed inside
the convection zone and radiated away in the polar regions not covered by
the stream. The effective temperature of these regions rises accordingly
but the stellar radius hardly changes. Numerical examples show that the
surface averaged temperatures of both components are close to one another
as observed in contact binaries but the calculated primary's temperature
cannot be lower than the secondary's. This is a known result; mass and
energy transfer between the components cannot by itself result in the
secondary's temperature higher than the primary's. The existence of A-type
binaries can be explained solely by large scale circulations but not
W-type. As has been suggested many times in the past, large dark
spots appearing temporarily on the primary can explain its low surface
averaged temperature. Numerical estimates obtained in the present paper
indicate that a substantial fraction of the star (25-50 percent) must be
covered with spots to lower the temperature by several hundred degrees,
required by observations.

\subsection{Discussion}

Dynamics of the mass transfer between the components of a contact binary,
considered in the present paper, is similar to the one investigated by
\citet{tas92}. The driving force of the circulations discussed by
\citet{tas92} comes also from the baroclinic structure of the common
envelope and the flow is influenced by the Coriolis force. The author
applied, however, a different boundary condition at the stellar
surface. Based on the observations available at that time he adopted zero
velocity and the strictly uniform temperature on the stellar surface of the
secondary. Yet more recent observations show the existence of massive flows
in some of the W UMa type stars \citep{pruc08}. On the other hand, very
little is known about the temperature distribution over the surfaces of
both components although nonuniformities in their surface brightness 
are notoriously invoked to explain the observed light curves
(e. g. \citealt{gaz06}).

\citet{ye05} called attention to a possible role of lateral energy transfer
connected with differential rotation observed on the Sun and solar-type
stars. Differential rotation results from a coupling between rotation and
turbulent convection in an axially symmetric star. AM is transported in the
meridional direction by Reynolds stresses, meridional circulations and
viscous diffusion \citep{bro08}. The mechanism considered in the present
paper is different from that. The differential rotation was not
included. Nevertheless, it may play a role in forming a stream which
transfers mass and energy between the components although more elaborated
model of the meridional AM transport in a contact binary is needed.

W UMa type stars show often period variations \citep{krei01}. Systematic
changes with values up to $\dot P = \pm (10^{-6} - 10^{-7})$ day/year have
been detected in several stars and interpreted as a signature of
TRO. However, recent observations of several hundred contact binaries
monitored in the framework of the program OGLE showed that the distribution
of period variations of W UMa stars can be approximated by a normal
distribution with an average equal to zero and dispersion equal to
$2.3\times 10^{-7}$ day/year \citep{kub06}. The observations are of very
high accuracy but they are extended only over 13 seasons.  The analysis of
the observed period variations suggest that they have a random character
with predominantly low values of $\dot P$ resulting in time scales
substantially longer than the stellar thermal time scale. Such a
distribution excludes TRO as a primary source of the variations. A much
more probable reason for the observed variations is connected with possible
fluctuations of the mass transfer between the components. The stream
flowing from the primary to the secondary and back carries $10^{-3} -
10^{-4} M_{\odot}$/year. A relative fluctuation of this flux at the level
of $10^{-3} - 10^{-4}$ can produce the observed period
variations. Fluctuations of that order are expected as a result of magnetic
activity cycles operating on W UMa stars \citep{app92}. The existence of
activity cycles have been suggested from the analysis of a long-time
photometric behavior of heavily spotted stars. The average magnitude of
those stars varies on time scales of several decades. This would be the
expected time scale of period variations in contact binaries.

The present model of large scale circulations supplements the evolutionary
model of W UMa type stars in which contact binaries are past mass exchange
with mass ratio reversal \citep{ste04, ste06a, ste06b} i. e. they are in an
evolutionary state similar to low mass Algols. The basic difference between
W UMa type stars and low mass Algols is that the latter stars have more
orbital AM and longer periods so only the secondary fills its Roche
lobe. Binaries with lower orbital AM form contact binaries in which the
overflow of the critical equipotential surface by both components drives
large scale circulations encircling the whole secondary and a large
fraction the primary.  Matter transported by the circulations to the
secondary (with the mass flux of the order of $10^{-3}-10^{-4}
M_{\odot}$/year) returns back to the primary but a fraction of thermal
energy associated with that matter is radiated away during the flow around
the secondary. It was argued that the expected flow velocities attain a
substantial fraction of the sound velocity. Such values were assumed when
considering details of the flow. To determine an accurate value of the flow
velocity numerical hydrodynamic simulations are needed. Particularly
important in this respect is the fate of the returning flow. If it behaves
like a waterfall, the expected flow velocities can be close to sound
velocity. If, instead, it collides with the primary's matter so that the
high pressure wave moves back along the flow, the driving force will be
reduced and the flow velocity lowered.

\section{Acknowledgments}
I thank Dr. Ryszard Sienkiewicz for calculating a set of evolutionary
models of low mass stars. The description of his program can be found in
\citet{pacz07}.  The remarks of an anonymous referee, which helped to
improve significantly the presentation of the present paper, are
highly appreciated.

\end{document}